# Approaching the knee – balloon-borne observations of cosmic ray composition

**Michael L Cherry**

Dept. of Physics and Astronomy, Louisiana State University, Baton Rouge, LA 70803 USA

E-mail: cherry@lsu.edu

**Abstract**. Below the knee in the cosmic ray spectrum, balloon and spacecraft experiments offer the capability of direct composition and energy measurements on the primary particles. A major difficulty is obtaining enough exposure to extend the range of direct measurements sufficiently high in energy to permit overlap with ground-based observations. Presently, balloon and space measurements extend only up to ~100 TeV, well below the range of ground-based experiments. The prospect of Ultra-Long Duration Balloon missions offers the promise of multiple long flights that can build up exposure. The status of balloon measurements to measure the high energy proton and nuclear composition and spectrum is reviewed, and the statistical considerations involved in searching for a steepening in the spectrum are discussed. Given the very steeply falling spectrum, it appears unlikely that balloon experiments will be able to extend the range of direct measurements beyond 1000 TeV any time in the near future. Especially given the recent suggestions from KASCADE that the proton spectrum steepens only at 4000-5000 TeV, the chance of detecting the knee with direct measurements of protons to iron on balloons is not likely to occur without significant increases in the payload and flight duration capabilities of high altitude balloons.

1. **Introduction**

The standard picture of cosmic ray acceleration to high energies is based on Fermi acceleration in expanding galactic supernova shocks. If protons are accelerated to a maximum energy $E = E_o$ and the acceleration is dependent on magnetic rigidity, then heavier nuclei of charge Z can be expected to be accelerated in the same supernova sources to energies $E_{max,Z} = ZE_o$. For acceleration at a parallel shock in a standard galactic supernova remnant, assuming a value of 3 μG for the interstellar field, Lagage and Cesarsky [1] calculated a maximum proton energy $E_o \sim 100$ TeV. This value can be extended upward with a number of mechanisms, for example by introducing higher fields, larger sources, quasi-perpendicular shocks, reacceleration by multiple sources, etc. Nevertheless, if there is a typical maximum energy which depends linearly on Z, then the spectrum of cosmic ray nuclei must become heavier with increasing energy as the hydrogen cuts off first and then increasingly heavier nuclei reach their acceleration limits. Since cosmic rays have been detected with energies much higher than 100 TeV, then either there is something drastically incorrect with the suggested supernova shock value $E_o$, or there must be a new source which turns on as the supernova shock mechanism reaches its maximum energy, as illustrated in Figure 1.

Particle-by-particle measurements on balloons and spacecraft have two enormous advantages over ground-based air shower experiments: 1) They detect the particle and measure its charge directly, and

2) spacecraft experiments perform the measurement above the entire atmosphere and balloon experiments typically perform their measurements with residual atmospheres ~ 5 g/cm$^2$, a small distance compared to typical interaction lengths. In order to probe energies approaching the "knee" near $10^{15}$-$10^{16}$ eV, where the all-particle spectrum is observed to steepen, however, the instruments to perform direct measurements must be large. This is especially true since current direct measurements have seen no evidence of the expected steepening in the spectrum out to energies approaching 100 TeV. If direct measurements are to be extended to 1000 TeV or higher, then a 1 m$^2$ sr calorimeter can be expected to detect only ~ 30 protons at E > 1000 TeV in a 900 day exposure. The maximum energy at which a spectral steepening can be confidently detected or ruled out is significantly lower than this.

There are several balloon-borne instruments currently developed or being developed with the capacity to measure the high energy cosmic ray spectrum. In Section 2, I review these instruments and their recent results, and in Section 3 I discuss the statistical issues involved in searching for the bend in the cosmic ray spectrum. In Section 4, I address the question: How large a detector will be required to extend current measurements upward in energy toward the knee region?

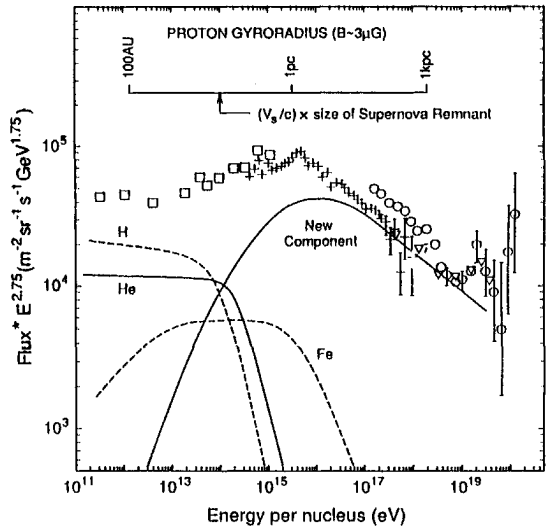

**Figure 1.** Sketch of the spectra expected for individual cosmic ray species in the standard supernova shock acceleration model [2]. The H, He, and Fe spectra are predicted to steepen near $10^{14}Z$ eV. The points show a sample of satellite and ground-based measurements of the all-particle spectrum. A schematic "new component" is needed to account for the observed flux above the expected galactic supernova cutoff.

2. **Current and recent balloon instruments to measure the high energy cosmic ray spectrum**
Table 1 lists seven current and recent instruments which either have provided recent results on the hydrogen - iron composition or are expecting to do so in the near future. JACEE and RUNJOB were passive emulsion/X-ray film calorimeters which flew ~11-15 flights each in order to accumulate the needed high energy statistics. Passive detectors have the advantage that the equipment is relatively inexpensive, and the detectors can be duplicated and flown multiple times in order to build up a large exposure factor. Multiple flights entail risks, however, that payloads may not be recovered: Both JACEE and RUNJOB experienced an unsuccessful campaign. Nevertheless, by combining the results from multiple campaigns, JACEE and RUNJOB have provided the highest energy direct measurements of the hydrogen - iron spectra that are currently available. JACEE has reported results based on a sample of 656 proton events above 6 TeV and 414 heliums above 2 TeV/nucleon, and RUNJOB has detected a proton event with energy near 1000 TeV. However, as shown below, these statistics make it possible to make statistically defensible statements about a steepening in the spectrum only out to ~40-90 TeV.

ATIC is a large (0.24 m² sr) electronic calorimeter with significantly greater depth than the emulsion instruments (10-22 radiation lengths compared to ~9 r.l. for JACEE). TRACER is the largest of the instruments: It is a $2 \times 2 \times 1.2$ m³ transition radiation detector with sensitivity from oxygen to iron. CREAM combines a calorimeter and a transition radiation detector. It is intended to fly on multiple very long flights, and has already accumulated 41 days of exposure in its initial Antarctic flight.

Finally, TIGER and CAKE are instruments designed to detect nuclei up to well above iron. Neither will provide data at extremely high energies, but they are included here because a complete picture of the cosmic ray composition requires extending current measurements both in energy and charge.

**Table 1.** Current and recent balloon instruments to measure high energy cosmic ray composition.

| Experiment | Description | # of balloon flights | | Reference |
|---|---|---|---|---|
| **JACEE** | Series of emulsion experiments | 11<br><br>1979-1994 | 644 m² hrs @ 3.5 – 5.5 g/cm²<br>Zenith angle acceptance out to tan θ ~<br>72 - 79° => ~80 m² sr days exposure | 3,4 |
| **RUNJOB** | Series of emulsion experiments | 10<br><br>1995-1999 | 575 m² hrs @ 9.0 – 10.7 g/cm²<br>Highest energy proton event seen at<br>E > 1 PeV | 5,6 |
| **ATIC** | Silicon matrix-scintillator-BGO calorimeter | 2<br><br>2000-2003 | 31 days flight time =><br>~ 7 m² sr days exposure | 7,8 |
| **TRACER** | Scintillator-Cerenkov-TRD for $8 \leq Z \leq 26$ | 2<br><br>1999-2004 | 15 days flight time =><br>~ 75 m² sr days exposure | 9-11 |
| **CREAM** | Scintillator-TRD-Si charge detector-W+scintillator calorimeter | 1<br><br>2004-2005 | 41 days flight time<br>Goal is to fly multiple 100 day ULDB flights to build up exposure | 12,13 |
| **TIGER** | Scintillator-Cerenkov-fiber hodoscope for $Z \geq 30$ | 3<br><br>1997-2004 | ~ 4 m² sr days exposure<br>Originally planned for ULDB flight;<br>future flights planned | 14 |
| **CAKE** | Nuclear track detectors (CR-39, Lexan) for $6 \leq Z \leq 74$ | 1<br><br>1999 | 22 hrs flight time @ 3-3.5 g/cm²<br>Planning to fly larger version on<br>ULDB | 15 |

2.1. *Japanese American Cooperative Emulsion Experiment (JACEE)*
The JACEE collaboration flew a series of thin (~8.5 radiation lengths) emulsion/X-ray film calorimeters on 15 flights, including 5 long duration Antarctic flights (JACEE 10-14). JACEE's published results on the hydrogen and helium spectra were based on data from 12 of those flights (Table 2). (JACEE 11 was lost at sea, and the data from JACEE 13-14 were never analyzed.) The binned data extended earlier results to energies close to 200 TeV for protons and ~100 TeV/nucleon for helium (Figure 2). In order to eliminate the uncertainty due to the arbitrary binning of data with small numbers of high energy events, the JACEE group presented integral spectra (Figure 3). A likelihood analysis gave best fit integral spectral indices for the protons and helium that differed at the 2σ level: $\gamma_H$ = 1.80 ± 0.04 and $\gamma_{He}$ = 1.68 ± 0.04/-0.06. If the difference in the spectral indices is real, it could indicate a different source for the hydrogen and helium. Searching for a steepening in a spectrum with meager statistics requires care, as will be described in some detail below in Section 3. The JACEE group reported no evidence of a steepening in their H spectra out to 40-90 TeV. Beyond that energy, the statistics allowed them to make no claims. Although the likelihood analysis was designed to eliminate correlations, the "waviness" observed in the integral spectra in Figure 3 raises an obvious concern: Is this an artifact of meager statistics or is it possibly an indication of systematic errors resulting from combining data from multiple flights with slightly different normalizations and calibrations? JACEE remains the balloon experiment with the highest statistics at energies above 10 TeV, but clearly <u>significantly</u> better statistics are needed to address the issue of the composition approaching the knee.

**Table 2.** JACEE balloon flights.

| JACEE Flight # | Launch date | Altitude (g/cm$^2$) | Duration (hrs) | # of emulsion blocks & individual size (cm x cm) | Cumulative exposure (m$^2$ hrs) |
|---|---|---|---|---|---|
| JACEE 0 | 5/79 | 8 | 29.0 | 1 (40 x 50) | 6 |
| JACEE 1 | 9/79 | 4 | 25.2 | 4 (40 x 50) | 26 |
| JACEE 2 | 10/80 | 4 | 29.6 | 4 (40 x 50) | 50 |
| JACEE 3 | 6/82 | 5 | 39.0 | 1 (50 x 50) | 59 |
| JACEE 4 | 9/83 | 5 | 59.5 | 4 (40 x 50) | 107 |
| JACEE 5 | 10/84 | 5 | 15.0 | 4 (40 x 50) | 119 |
| JACEE 6 | 5/86 | 4 | 30.0 | 4 (40 x 50) | 143 |
| JACEE 7 | 1/87 | 6 | 150.0 | 3 (40 x 50) | 233 |
| JACEE 8 | 2/88 | 5 | 120.0 | 3 (40 x 50) | 305 |
| JACEE 9 | 10/90 | 4 | 44.0 | 4 (40 x 50) | 340 |
| JACEE 10 | 12/90 | 4 | 204.0 | 2 (30 x 40) | 389 |
| JACEE 12 | 1/94 | 5 | 212.0 | 6 (40 x 50) | 644 |

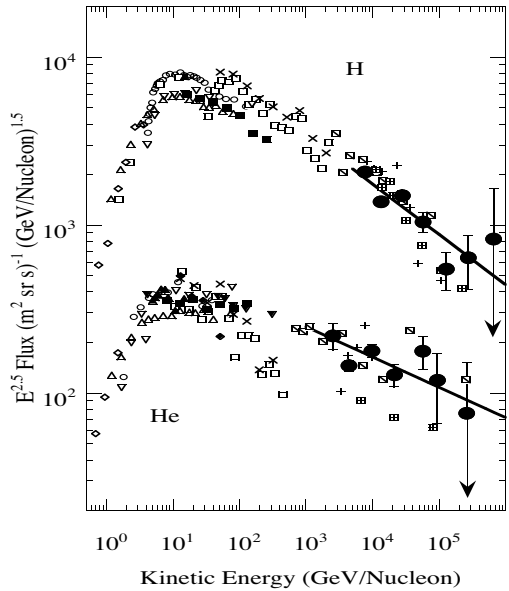

**Figure 2.** JACEE 1-12 differential spectra for H and He together with earlier results [3].

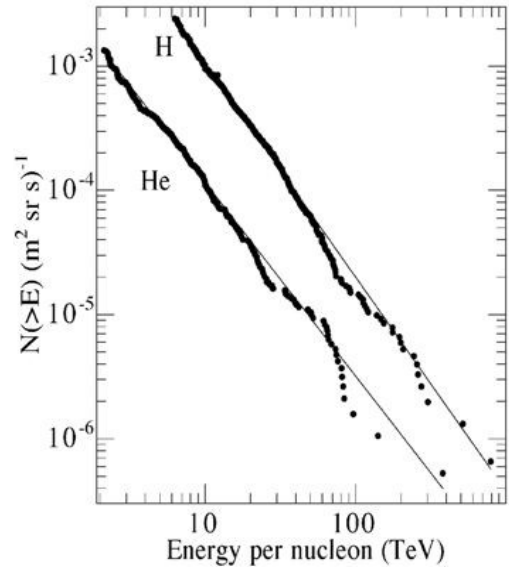

**Figure 3.** JACEE 1-12 integral spectra for hydrogen and helium.

*2.2 RUssia-Nippon JOint Balloon collaboration (RUNJOB)*

RUNJOB flew a roughly similar set of X-ray films and emulsion chambers on a series of 11 balloon flights from Kamchatka starting in 1995, of which 10 were successful with a total of 575 $m^2$ sr of exposure. Although RUNJOB flew somewhat lower in the atmosphere (~10 $g/cm^2$) than did JACEE, RUNJOB has presented results for hydrogen through iron. The integral spectral indices for the RUNJOB hydrogen and helium are equal to within the statistical errors: $\gamma_H = 1.74 \pm 0.08$ and $\gamma_{He} = 1.78 \pm 0.20$ at energies up to 100 TeV. In the 1995 campaign, RUNJOB detected a single proton with energy near 1000 TeV and, over 10 flights, saw no evidence for any steepening in the proton spectrum with increasing energy. The absolute intensity of the hydrogen measured by RUNJOB agreed reasonably well with the intensity measured by JACEE, but RUNJOB reported a helium intensity in agreement with the intensity reported by MUBEE and roughly half that of JACEE and SOKOL.

For charges heavier than helium, RUNJOB presented the energy spectra of three charge groups: CNO, NeMgSi, and Fe, with a slope reported to harden from ~1.7 for CNO to ~1.6 for Fe. The resulting average mass <ln A> reported by RUNJOB was roughly constant at ~1.5 up to 1000 TeV/nucleus, in reasonable agreement with the results from the KASCADE air shower experiment and the BLANCA Cerenkov light measurement but in disagreement with the results from CASA-MIA, which showed <ln A> increasing with energy starting well below 1000 TeV/nucleus.

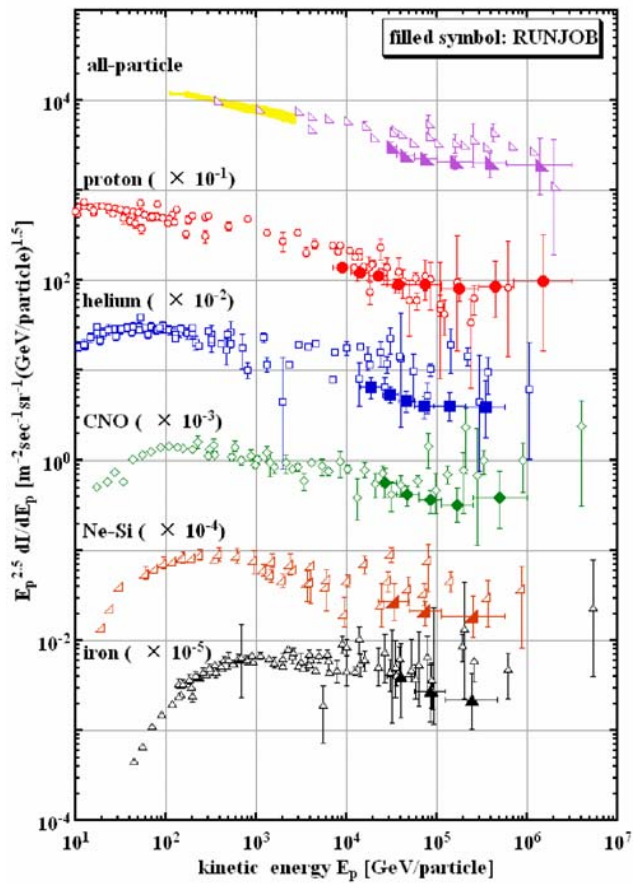

**Figure 4.** Final hydrogen - iron spectra obtained by RUNJOB (solid points) compared to other measurements and the all-particle spectrum [6]. In particular, note that the absolute proton intensities reported by RUNJOB (●) and JACEE (○) are in reasonable agreement while the helium intensities (■ and □ respectively) differ by approximately a factor of 2.

*2.3 Advanced Thin Ionization Calorimeter (ATIC)*

ATIC is an active electronic instrument combining a fine-grained silicon matrix charge detector and an 18 radiation length deep (22 r.l. for ATIC-3) bismuth germanate (BGO) calorimeter with a geometry factor of 0.24 $m^2$ sr. ATIC is intended to provide sensitivity for hydrogen through nickel from 50 GeV to ~100 TeV total energy. ATIC has flown twice from the Antarctic with a combined 31 days of exposure, and is scheduled for a third flight in December 2005. The preliminary ATIC-2 results show reasonably flat spectra for both hydrogen and helium (Figure 5) with spectral indices $\gamma_H \sim 1.9$ and $\gamma_{He}$

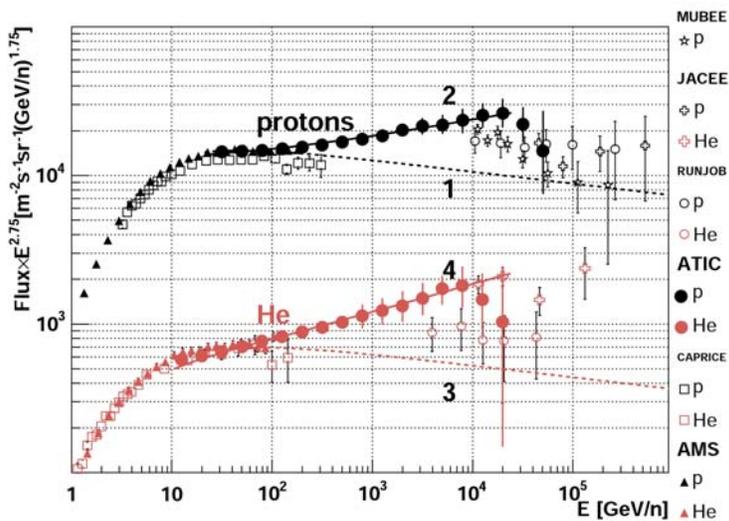

**Figure 5.** Proton and helium spectra measured in ATIC-2 compared to earlier measurements [8]. Lines correspond to predictions from Leaky Box model (1 and 3) and diffusion model with reacceleration (2 and 4).

~ 1.83, in good agreement with predictions from a Leaky Box diffusion model with weak reacceleration. (ATIC-1 showed somewhat flatter spectra, essentially identical for hydrogen and helium. The differences appear to be due to better statistics and an improved procedure for fitting the ATIC-2 spectra.) In the case of the heavier nuclei, ATIC has sufficient charge resolution to identify individual charges. This is in contrast to the emulsion instruments, which showed nuclear results for charge groups only. The ATIC-2 nuclear spectral results are shown in Figure 6, where the statistics are clearly not yet sufficient to be able to make strong claims at energies above 1 TeV/nucleon.

*2.4 Transition Radiation Array for Cosmic Energetic Radiation (TRACER)*

TRACER is the largest of the current generation of balloon-borne detectors. It utilizes a 2 m × 2 m × 1.2 m high array of transition radiation detectors together with a set of plastic scintillators and an acrylic Cherenkov counter to measure the charge and energies of nuclei with $Z \geq 8$. The relatively low density of a transition radiation detector utilizing plastic fiber radiators and multiwire proportional chambers makes it possible to design the instrument with the very large collecting power (~ 5 m² sr) needed to perform a measurement of the nuclear composition at energies of 100 – 1000 TeV/nucleus with reasonable statistical significance. TRACER has flown twice – a 30 hour test flight from Ft.

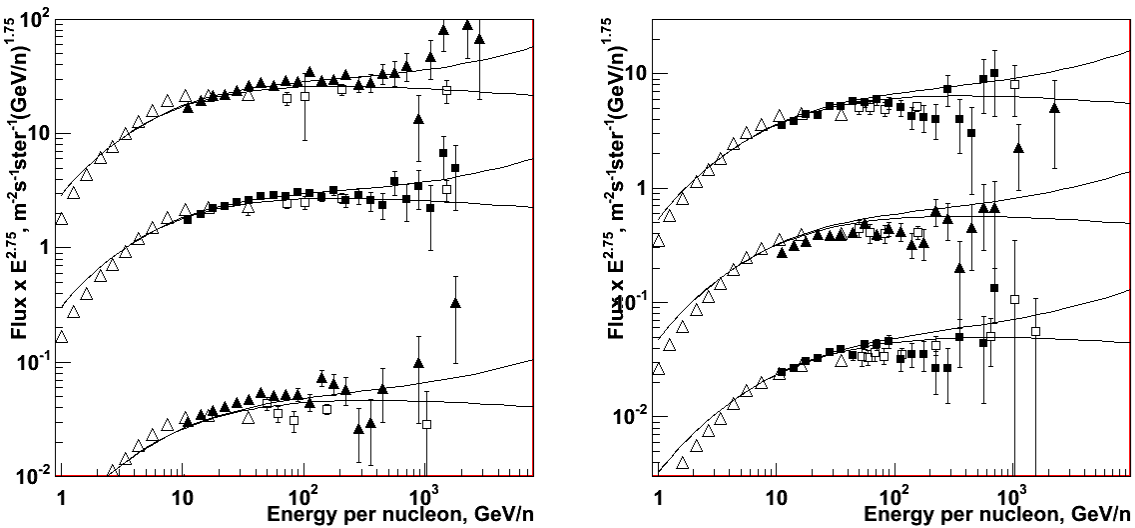

**Figure 6.** Energy spectra of abundant nuclei from ATIC-2 (▲ and ■). Left hand side: Data from top to bottom are for C, O/10, and Ne/100. Right hand side: Data from top to bottom are for Mg, Si/10, and Fe/100. Also shown are results from HEAO-3-C2 (Δ) and CRN (□).

Sumner and a 14 day exposure from the Antarctic. Absolute intensities and the Fe, Si, and Mg spectral slopes measured on the initial flight agree well with previous observations in space with CRN and HEAO, although the preliminary TRACER Ne and O spectra may be slightly flatter than those reported from CRN. More details about the most recent results are given in [11].

*2.5 Cosmic Ray Energetics And Mass (CREAM).*
CREAM combines a segmented silicon charge detector, a scintillating fiber hodoscope, a pair of 120 × 120 × 35 cm$^3$ transition radiators, and a 20 radiation length 50 × 50 cm$^2$ tungsten-scintillator sampling calorimeter. The calorimeter provides sensitivity for hydrogen and heavier particles, and the transition radiation detector is designed to be used for Be to Ni. The geometry factor for protons entering the top of the 0.5 $\lambda_{int}$ carbon target and passing through at least 20 radiation lengths in the calorimeter is 0.35 m$^2$ sr; for heavy nuclei (with a shorter interaction mean free path in the target), the calorimeter's effective geometry factor increases to 0.57 – 0.7 m$^2$ sr; and the transition radiation detectors provide an additional 1.3 m$^2$ sr. The instrument has been designed for 100-day Ultra Long Duration Balloon (ULDB) flights, with a goal of flying multiple times in order to obtain statistically significant spectral and composition information for protons through iron from 10$^{12}$ to 10$^{15}$ eV.

CREAM had an initial Antarctic flight in December 2004 – January 2005 which set a record with nearly 42 days exposure at an altitude of ~120,000 – 130,000 ft. Assuming 100% livetime and acceptance, the 0.35 m$^2$ sr calorimeter geometry factor translates to an exposure of nearly 15 m$^2$ sr days in just the initial flight, already on the order of 10% of the combined JACEE-RUNJOB exposure. CREAM is scheduled to fly a second time in December 2005.

*2.6. Experiments to measure the abundances of Z>26 nuclei.*
Two current instruments are studying the composition of heavy (Z > 26) nuclei: TIGER (Trans-Iron Galactic Element Recorder) and CAKE (Cosmic ray Abundances below Knee Energy). TIGER uses a combination of scintillators together with plastic and Aerogel Cherenkov detectors and a scintillating fiber hodoscope to measure the charge and energy of incident nuclei for charges from Z ~ 26 to 40 at energies up to approximately 10 GeV/nucleon. TIGER was the first ULDB payload selected, but because of delays in the ULDB program, it has flown three times on shorter flights – once for a day from Ft. Sumner and twice from the Antarctic for 50 days of additional exposure. Measured abundances appear to be in good agreement with solar system abundances modified by either first ionization potential (FIP) or volatility, but the statistics are not yet good enough to distinguish between FIP and volatility. CAKE employs stacks of CR39 and Lexan nuclear track detectors to identify charges in the range 6 < Z < 74 with a geometry factor of 1 – 2 m$^2$ sr. An initial 22 hour test flight from Italy to Spain was conducted in 1999, and the goal is to fly a larger version of the instrument on a ULDB campaign.

**3. Statistical analysis of a spectral break: Where is the bend in the high energy spectrum?**
To fit low-statistics spectra and search for statistical evidence of a break, the JACEE group used a Poisson-weighted maximum likelihood approach [4]. As discussed above, binning low-statistics data introduces a possibility of systematic errors depending on the exact (arbitrary) choice of bin boundaries. To eliminate this effect, JACEE worked with the individual events plotted as an integral spectrum in Figure 3. The n individual events are ordered by decreasing energy from $E_1 = E_{max}$ to $E_n = E_{min}$. In any interval $E_{i-1} - E_i$, the expected number of events is

$$< n >_i = dN/dE_i \, (E_{i-1} - E_i) \, (G/\varepsilon)_i \qquad (1)$$

where $G_i$ is the geometry (acceptance) factor (m$^2$-sr-s-TeV) for the i-th interval and $\varepsilon_i$ is the expected efficiency. The Poisson probability of seeing one event when $<n>_i$ are expected is then

$$P_i(n=1) = <n>_i \, e^{-<n>_i} \, . \qquad (2)$$

If a broken power law is assumed of the form

$$dN/dE = aE^{-(\gamma+1)} / [1 + (E/E_o)^\delta] \quad , \tag{3}$$

one can evaluate the likelihood of the spectrum (3) from the product of the probabilities

$$\ln L = \ln \Pi_i P_i = \Sigma_i \ln \langle n \rangle_i - N_{events} \tag{4}$$

where $N_{events}$ is the total number of points in the spectrum. The fitted power law spectra are obtained by setting $E_0$ to be large and varying the parameters a and $\gamma$ in Equation (3) in order to maximize ln L.

In order to look for evidence of a break in the spectrum near energy $E_0$ (i.e., a steepening from a low-energy spectrum with power law index $\gamma_1$ to a high-energy spectrum with index $\gamma_2$), one can constrain the low-energy spectral index $\gamma_1 = \gamma$ to be in the range $1.6 \leq \gamma_1 \leq 1.7$ based on the measured low-energy data. At high energies, the limiting high-energy index $\gamma_2 = \gamma + \delta$ is expected to be in the range $2.0 \leq \gamma_2 \leq 2.3$. By requiring a steepening somewhere in the range 1 - $10^4$ TeV, $\delta$ is constrained to be in the range $0.3 \leq \delta \leq 0.7$. As a function of the break energy $E_0$, and for values of $\delta$ = 0.3, 0.4, 0.5, 0.6, and 0.7, one then chooses a and $\gamma$ to maximize -ln L and plots the maximized -ln L vs. $E_0$. Figure 7 shows the result for the JACEE proton data. Also in Figure 7, we show the maximized likelihood for the case of no break ($\delta$ = 0). In all cases where a break is allowed ($\delta$ > 0), the maximized likelihood peaks gradually near 100 TeV. The difference between the peak value of -ln L (approximately 613.5 to 614.2) and the value without a break (614.7) is small (~0.1%), and the peak occurs at an energy sufficiently high that it is governed mainly by the ~20 protons above 100 TeV. In other words, within 0.1% in -ln L, it is equally likely that there is or is not a break in the spectrum.

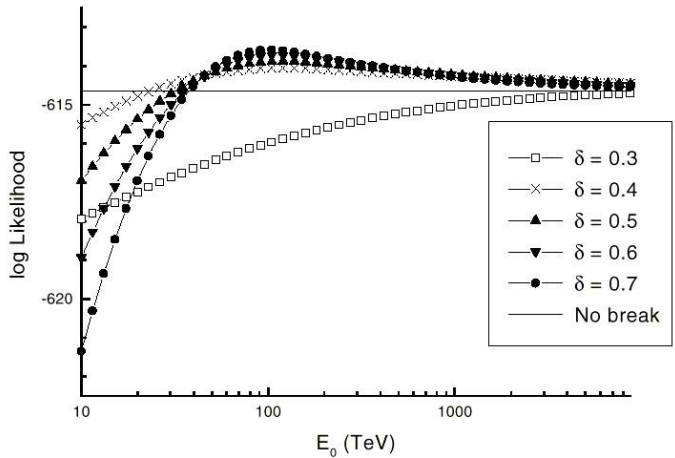

**Figure 7.** Log likelihood of a steepening in the JACEE proton spectrum by $\delta$ in the power law spectral index at a break energy $E_0$

The maximum energy at which an experiment has sufficient statistical accuracy to detect or rule out a bend in the spectrum can be evaluated as follows: In the case of the JACEE protons, if the measured events are taken to be a parent distribution, randomized data sets can be generated by applying Poisson statistics to the parent set and -ln L can be maximized for the randomized data. The random statistical variation in ln L (for the assumed case of no break in the spectrum, $\delta$ = 0) is then $\sim \pm 25$, much larger than the difference between the curves with and without a break in Figure 7. This observed variation of $\pm 25$ is expected from Equation (4): In the case of an unlikely but perfect fit ($\langle n \rangle_i = 1$ for all i), -ln L = $N_{events}$ and therefore has a standard deviation (due to statistics only, and for $N_{events}$ sufficiently large) equal to $\sqrt{N_{events}}$. This is ~25 for the JACEE proton data. In Figure 7, a break in the spectrum corresponds to the ratio

$$R = \frac{L_{break}}{L_{no\,break}} = \frac{\Pi_i P_{i,break}}{\Pi_i P_{i,no\,break}} = \frac{\Pi_i \langle n \rangle_i^b e^{-N_{events}}}{\Pi_i \langle n \rangle_i^n e^{-N_{events}}} \tag{5}$$

in excess of unity. The numerator in Equation (5) is calculated with a broken power law spectrum of the form (3); the denominator is calculated with a straight spectrum where $E_o \rightarrow \infty$. (We note that in both the numerator and the denominator, the fitted spectrum is normalized so that $\Sigma_i <n>_i = N_{events}$.) Given the uncertainty derived above for ln L, and the corresponding uncertainty in ln R, the condition that a break occurs in the spectrum corresponds to the requirement that (at 68% confidence)

$$\ln R \geq \sqrt{N_{events}} \quad . \tag{6}$$

In order to satisfy (6), it can be shown that the number of events at energies above $E_o$ must satisfy

$$N_{>E_o} \geq \xi^{-1} \sqrt{N_{events}} \sim (1-3)\sqrt{N_{events}} \tag{7}$$

where $\xi$ is a numerical factor approximately equal to the value of $\ln[1+(E/E_o)^\delta]$ averaged over all points i with $E_i > E_o$. Since the energies corresponding to 25 and 75 events in the JACEE proton data set are at 90 and 40 TeV respectively, below the peaks in Figure 7, we conclude that JACEE has no evidence for or against any bend in the high energy spectrum up to 40 – 90 TeV.

This argument can now be used to determine the maximum energy at which any experiment has sufficient statistical accuracy to detect a bend in the proton spectrum. In the case of JACEE, this corresponds to a maximum detectable $E_o \sim 40 - 90$ TeV. For an experiment with 50 times the collecting power (geometry factor × efficiency × exposure time), $N_{>E_o}$ increases by ~7 based on Equation 7. The original JACEE, however, would see 50 times fewer events than the new (50 times larger) experiment, and so $E_o$ corresponds to the point where JACEE would see (25-75)/7 ~ 4-10 events. In Figure 3, this would correspond to an energy of 200-300 TeV depending on the value of $\delta$.

## 4. Discussion and conclusions

Photon observations across the wavelength band show that the high energy sky is filled with transient, violent sources of high energy particle acceleration on scales from solar flares to active galactic nuclei. Gamma ray bursts and galactic microquasars are examples of explosive activity accompanied by relativistic jets which are undoubtedly sources of energetic particles. If the cosmic ray accelerator is a complex composite of young stars, stellar winds, supernovae, pulsars, jets, shocks in star forming regions, and various other sources, then a distinct cutoff in the spectrum at $Z \times 10^{14}$ eV from a single standard supernova source type transforms into a sum over an unknown distribution of magnetic field strengths, acceleration region sizes, shock strengths, stellar masses, etc. A "multiple source type" model could lead to a much slower variation of composition with energy than the standard "single source type" supernova model together with higher $E_{max}$ values. A definitive experiment must then be capable of working at energies significantly higher than JACEE's current 40 – 90 TeV limit on $E_o$ and must be sensitive to relatively modest steepening in the spectrum (e.g., $\delta \sim 0.3 - 0.7$ as discussed in Section 3 above) as opposed to a sharp cutoff.

The required size of a detector can be estimated as follows: For H and He, consider a calorimeter of 0.9 m$^2$ sr (the size used in one of the recent designs for a calorimeter for the ACCESS mission [16] and 3 – 4 times the size of the ATIC and CREAM calorimeters). For B-Fe, assume a transition radiator with geometry factor 6 m$^2$ sr (somewhat larger than TRACER). One can estimate the rates in such a large detector by taking the hydrogen spectrum to be a straight power law with integral spectral index $\gamma_H = 1.75$ and no break. The spectral index for all other nuclei is assumed to be $\gamma_Z = 1.65$ except for B which has a steeper spectrum: $\gamma_B = 2.1$. For this assumed model, the numbers of events expected on a 100-day balloon flight are given in Table 3. Even for this very large assumed detector, this very reasonable model spectrum leads to very few events expected above a total energy of 1000 TeV. Given the size of the detectors assumed here, one might want to retreat from the ACCESS-CREAM approach of combining multiple detectors in a single mission. Rather, one might choose to emphasize either the p-He measurement with a calorimeter, B-CNO with a TRD, or C-Fe with a TRD. Even with these limited goals, the detectors must be very large and the exposure times must be very long.

**Table 3.** Expected numbers of events in a 100 day Ultra Long Duration Balloon flight with a 0.9 m$^2$ sr calorimeter, 6 m$^2$ sr transition radiation detector, and spectrum as described in the text. Adapted from ref. [16].

| E (TeV) | H | He | B | C | Si | Fe |
|---------|---|----|---|---|----|----|
| 0.3 – 1 | $3.2 \times 10^6$ | $2.3 \times 10^6$ | $1.6 \times 10^4$ | $1.7 \times 10^5$ | $1.5 \times 10^5$ | $3.6 \times 10^5$ |
| 1 - 3 | $3.8 \times 10^5$ | $3.1 \times 10^5$ | $1.5 \times 10^3$ | $2.3 \times 10^4$ | $1.9 \times 10^4$ | $4.8 \times 10^4$ |
| 3 – 10 | $5.7 \times 10^4$ | $5.2 \times 10^4$ | 180 | $3.8 \times 10^3$ | $3.2 \times 10^3$ | $8.0 \times 10^3$ |
| 10 – 50 | $7.5 \times 10^3$ | $7.7 \times 10^3$ | 18 | 550 | 470 | $1.2 \times 10^3$ |
| > 100 | 140 | 190 | 3 | 20 | 160 | 410 |
| > 300 | 21 | 30 | | 4 | 4 | 67 |
| > 1000 | 3 | 4 | | | | 9 |

NASA plans for its new 100-day ULDB program currently call for a payload capacity of 1 ton [17]. ATIC, CREAM, TRACER, TIGER, and CAKE are all in excess of this proposed limit. Assuming that launch limits can eventually be raised sufficiently to allow the kinds of very large payloads envisioned in Table 3, then a series of 40 100-day flights would amount to 3600 m$^2$ sr days of calorimeter exposure, roughly 50 times the accumulated JACEE hydrogen-helium exposure. As shown in Section 3 above, an exposure 50 times that of JACEE would enable one to extend a definitive measurement of $E_o$ up to 200 – 300 TeV. Even the aggressive CREAM program, which envisions multiple flights over many years and which achieved ~10% of the JACEE-RUNJOB exposure in its initial flight, will require a decade of 40-day flights every year just to match the accumulated JACEE-RUNJOB exposure. Given foreseeable balloon capabilities, it is hard to envision extending direct measurements higher in energy than ~ 200 TeV with balloons.